\documentclass[aps,prl,reprint,twocolumn,superscriptaddress,showpacs,showkeys,groupedaddress]{revtex4-1}

\usepackage{amsmath,amssymb,amstext}
\usepackage[usenames,dvipsnames]{color}
\usepackage{graphicx}
\usepackage{bm,bbold,braket}
\usepackage{natbib}
\usepackage{txfonts, comment,stmaryrd}
\usepackage{dcolumn}
\usepackage[english]{babel}

\usepackage[colorlinks,bookmarks=false,citecolor=blue,linkcolor=red,urlcolor=blue]{hyperref}
\begin{document}

\title{Periodically Driven Array of Single Rydberg Atoms}

\author{Sagarika Basak, Yashwant Chougale and Rejish Nath}

\affiliation{Indian Institute of Science Education and Research, Pune 411 008, India}    

\begin{abstract}
An array of single Rydberg atoms driven by a frequency modulated light field is studied. The periodic modulation effectively modifies the Rabi coupling, leading to unprecedented dynamics in the presence of Rydberg-Rydberg interactions. They include state dependent population trapping, the Rydberg blockade for small and anti-blockades at large interaction strengths. Interestingly, the Schrieffer-wolf transformation reveals a fundamental process in Rydberg gases, correlated Rabi oscillations, arising from the long-range interactions, provides an alternative depiction for Rydberg blockade and it exhibits a nontrivial behaviour in the presence of periodic modulation. The dynamical localization of a many body configuration in a driven Rydberg-lattice is discussed.
\end{abstract}



\maketitle

Controlled coherent quantum dynamics is a great challenge and provides the impetus for various studies in diverse systems; in particular, periodically driven or Floquet  systems \cite{flo,she10}. They exhibit a wealth of quantum phenomena, in single particle case, the noted ones are the coherent destruction of tunneling \cite{gro91}, dynamic localization in quantum transport \cite{rag96}, population trapping (PT) in a two-level system (TLS) \cite{2lev,noe98} and Landau-Zener-St\"uckelberg (LZS) interference \cite{she10,lzs, sun09}. The latter happens when the TLS is driven periodically across an avoided crossing such that separate Landau-Zener transitions (LZTs) interfere \cite{she10, bec}, also demonstrated using Rydberg atoms \cite{noe98,ryd}.  The quantum interference plays a decisive role in above studies. Extending to many-body systems, complex scenarios emerge, for instance, dynamical-freezing \cite{fre},  -synchronization \cite{syn}, and -localization \cite{loc}, as  shown in driven quantum spin systems. Also, ergodicity breaking or many body localization in driven quantum systems has been the subject of intense theoretical studies \cite{mbl}, and recently, probed experimentally in a lattice of interacting ultracold fermions \cite{mblex}. On a bigger perspective, they constitute a platform to explore non-equilibrium quantum states \cite{none} and topological phenomena \cite{top}, including the time crystals \cite{tcry}.

On the other side, ultra-cold Rydberg gases \cite{ryd} offer a manifold of prospects to probe quantum physics  \cite{rqu}, owing to their exaggerated properties \cite{ryd}, supported by the experimental developments \cite{rex}; in particular, recent realizations of Rydberg-atomic arrays \cite{rch, zei17}. A Rydberg setup typically involves laser fields coupling the ground state $|g\rangle$ with one or more Rydberg states $|e\rangle$. At very low temperatures, the strong interactions among the Rydberg atoms lead us to describe the internal-states dynamics in the frozen gas limit, where the motional degrees of freedom are neglected \cite{fro}. The level shifts caused by the interactions suppress further excitations within a finite volume is called the Rydberg blockade \cite{blo,blo2}.  The latter brings up the {\em super-atom} picture, in which a fully blockaded ensemble of $N$ atoms exhibits Rabi oscillations between the ground state $|G\rangle=\otimes_{i=1}^{N}|g^{(i)}\rangle$ and the collective single excitation $|+\rangle=\sum_i|g g . . . e^{(i)} . . . g g\rangle/\sqrt{N}$ \cite{rab}. It has been proposed to generate entangled mesoscopic ensembles for fast quantum gate operations \cite{blo}.

In this letter, we examine a chain of single atoms driven by a frequency modulated field, which couples $|g\rangle$ to $|e\rangle$. The periodic modulation effectively modifies the Rabi couplings and together with Rydberg-Rydberg interactions unprecedented scenarios emerge. For instance, the Rydberg blockade exists even for interactions weaker compared to single atom Rabi coupling, resonant excitation of $|ee\rangle$ at large interactions (anti-blockade), and state dependent PT. The blockade enhancement offers the possibility of entangling two atoms at large separations without altering the bare Rabi frequency. The anti-blockade in a non-driven setup demands a three-level scheme \cite{abl} or a zero-area phase jump pulse \cite{abl2}, but the population in $|ee\rangle$ is found to be very small \cite{abl}, which can be significantly augmented in a driven system. Interestingly, employing the Schrieffer-wolf transformation \cite{swt} reveals a qualitatively novel feature, {\em correlated} Rabi-oscillations, arising from the long-range nature of the interactions, analogous to density assisted hopping in optical lattices \cite{das}. Correlated Rabi coupling (CRC) provides an alternative depiction for Rydberg blockade. Finally, we discuss the interaction dependent dynamical localization of a many-body configuration, which may pave a way towards exploring ergodic-nonergodic transitions using periodically driven Rydberg chains.

{\em Model}. We consider an one dimensional array with one atom per site, in which the electronic ground state $|g\rangle$ is coupled to a Rydberg state $|e\rangle$ via a light field with its frequency modulated periodically in time $t$. The system is described in the frozen gas limit, by the Hamiltonian:
\begin{equation}
\hat H=-\hbar\Delta(t)\sum_{i=1}^N\hat\sigma_{ee}^{i}+\frac{\hbar\Omega}{2}\sum_{i=1}^N\hat\sigma_x^{i}+\sum_{i<j}V_{ij}\hat\sigma_{ee}^{i}\hat\sigma_{ee}^{j},
\label{ham}
\end{equation}
where $\hat\sigma_{ab}=|a\rangle\langle b|$ with $a, b\in \{e, g\}$, $\hat\sigma_x=\hat\sigma_{eg}+\hat\sigma_{ge}$, $\Omega$ is the Rabi frequency, $\Delta(t)=\Delta_0+\delta\sin\omega_0t$ is the time-dependent detuning with amplitude $\delta>0$ and the modulation frequency $\omega_0$. An alternative way to introduce periodic $\Delta(t)$ is to drive the Rydberg state using a modulated microwave field \cite{sun09}, which off-resonantly couples to a nearby Rydberg state \cite{noe98}. The Rydberg excited atoms interact via the strong van der Waals interactions, $V(r)=C_6/r^6$ \cite{vwi}. We solve numerically the Schr\"odinger equation: $i\hbar\partial/\partial t|\psi\rangle=\hat H(t)|\psi\rangle$ and analyze the dynamics. Henceforth we take $\hbar=1$.

To gain an insight, especially at large $\omega_0$ or interactions, we move to a rotating frame  \cite{swt}: $|\psi'\rangle=\hat U(t)|\psi\rangle$ where $\hat U(t)=\exp[if(t)\sum_j\hat\sigma_{ee}^{j}+it\sum_{j<k}V_{jk}\hat\sigma_{ee}^{j}\hat\sigma_{ee}^{k}]$ with $f(t)=\delta/\omega_0\cos\omega_0 t-\Delta_0 t$.  The new Hamiltonian, $\hat H'(t)=\hat U\hat H\hat U^{\dagger}-i\hbar\hat U\dot{\hat U}^{\dagger}$, after using the Jacobi-Anger expansion $\exp({\pm iz\cos\omega_0 t})=\sum_{m=-\infty}^{\infty}J_m(z)\exp(\pm im[\omega_0t+\pi/2])$, is \cite{supp}
\begin{eqnarray}
\hat H'&=&\frac{\hbar\Omega}{2}\sum_{j=1}^N\sum_{m=-\infty}^{\infty}i^mJ_m(\alpha)\left(g(t)e^{i\sum_{k\neq j}V_{jk}\hat\sigma_{ee}^{k}t}\hat\sigma_{eg}^{j}+{\rm H.c.}\right)  
\label{hamro}
\end{eqnarray}
where $J_m(\alpha)$ is the $m$th order Bessel function with $\alpha=\delta/\omega_0$ and $g(t)=\exp[i(m\omega_0-\Delta_0)t]$. As seen in Eq. (\ref{hamro}), the periodic detuning has effectively modified the Rabi coupling. 


{\em Single atom}. Hinge on $\Omega$, $\delta$ and $\omega_0$, we consider three regimes \cite{2lev,ash}. (i) The weak driving limit: $\delta\ll \Omega_{eff}\equiv\sqrt{\Omega^2+\Delta_0^2}$, the resonant $|g\rangle$ to $|e\rangle$ transition occurs at $\omega_0=\Omega_{eff}$ with an effective Rabi frequency $\Omega'=\delta\sin({\rm tan}^{-1}[\Omega/\Delta_0])$ \cite{ash}. (ii) The high-frequency limit (HFL): $\omega_0\gg\Omega$, the only term relevant in $\sum_{m=-\infty}^\infty(\cdot)$ in Eq. (\ref{hamro}) at longer times gives the resonance $m\omega_0=\Delta_0$, with $\Omega'\approx\Omega |J_m(\alpha)|$. If $\Delta_0=0$, $\Omega'=\Omega J_0(\alpha)$, hence at $J_0(\alpha)=0$, the PT happens \cite{2lev,noe98}. (iii) The fast passage limit (FPL): $\omega_0\sqrt{\delta^2-\Delta_0^2}\gg\Omega^2$ with $\delta-\Delta_0\gg\Omega$, the resonance is at $n\omega_0=\Delta_0$ with $\Omega'=(2\omega_0/\pi)|\cos(\theta-\pi/4)|\sqrt{\frac{\pi\Omega^2}{2\omega_0}/\sqrt{\delta^2-\Delta_0^2}}$, where $\omega_0\theta=\sqrt{\delta^2-\Delta_0^2}-\Delta_0\cos^{-1}\frac{\Delta_0}{\delta}$ \cite{ash}. The PT occurs if $\cos(\cdot)=0$. Both HFL and FPL co-exist for $\delta, \omega_0\gg \Omega$. In the FPL, the system is driven past an avoided level crossing repeatedly, causing non-adiabatic LZTs. The transfer matrix (TM) method based on adiabatic-impulse model  \cite{she10,ash} gives a good description. Around the avoided crossing, the LZTs are given by a non-adiabatic unitary operator $\hat G_{LZ}$ and away from it, an adiabatic phase evolution through $\hat G_j$ with $j$ being left ($j=1$) or right ($j=2$) of the crossing. Then, the total phase acquired during a  full cycle provides the resonant conditions.   

\begin{figure}[hbt]
\centering
\includegraphics[width= 1.\columnwidth]{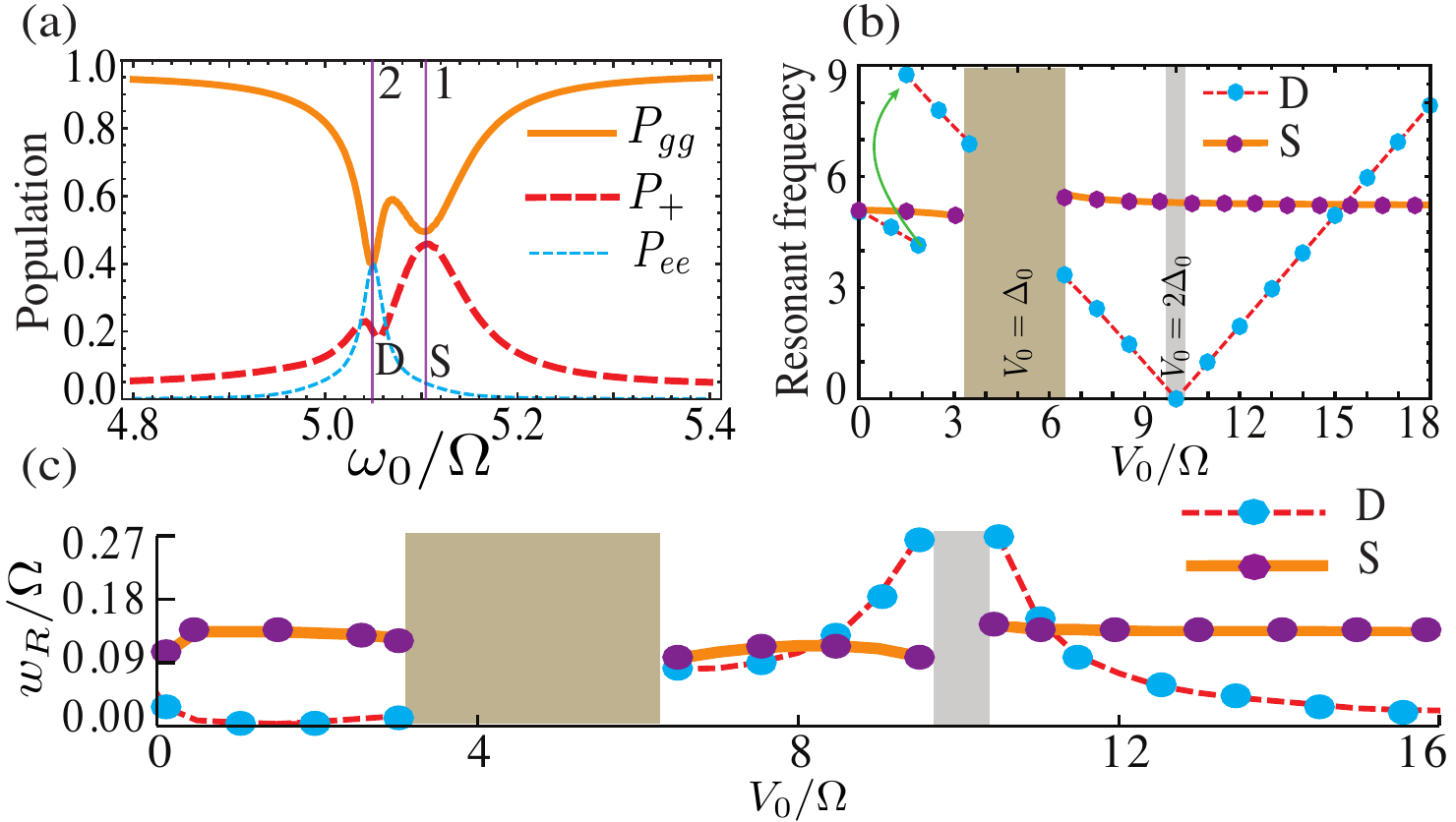}
\caption{\small{(color online). The results for $N=2$ in the weak driving limit with $\delta=0.4\Omega$ and $\Delta_0=5\Omega$. (a) $P_{\beta}(t)$ vs $\omega_0$ with $|\beta\rangle\in\{|gg\rangle, |+\rangle, |ee\rangle\}$ for $V_0=0.1\Omega$ and $\Omega T_f=1000$. The vertical lines 1 and 2  indicate resonances $S$ and $D$ respectively. (b) $\omega_S/\Omega$ (solid lines) and $\omega_D/\Omega$ (dashed lines) vs $V_0$ and the numerical results are shown by filled circles. (c) the widths, $w_R$ of resonances vs $V_0$.}}
\label{fig:1} 
\end{figure}
{\em $N=2$}. The interactions, $V(r_{12})=C_6/a^6\equiv V_0$, where $a$ is the lattice spacing, significantly modifies the resonance criteria as well as the excitation dynamics. In particular, we focus at the resonance criteria for $|gg\rangle\leftrightarrow |+\rangle=(|eg\rangle+|ge\rangle)/\sqrt{2}$ ($S$) and $|gg\rangle\leftrightarrow |ee\rangle$ ($D$) transitions. $S$ and $D$ indicate single and double excitations respectively. Numerically, they are obtained as the peaks/dips in the time averaged populations: $P_{\beta}=(1/T_f)\int_0^{T_f}|\langle \beta|\psi(t)\rangle|^2dt$; $|\beta\rangle \in\{|gg\rangle, |ee\rangle, |+\rangle\}$, as a function of $\omega_0$, with an initial state $|I\rangle=|gg\rangle$ [see Fig.\ref{fig:1}(a) and (b)]. 

{\em Weak driving limit}. For $V_0\ll\Omega_{eff}$, [Fig.\ref{fig:1}(a)], the atoms are assumed non-interacting, and $V_0$ is account through an effective detuning, $\Delta_0'=\Delta_0-V_0/2$ for $|ee\rangle$. Thus, the resonances occur at $\omega_0=\omega_S=\Omega_{eff}$ and $\omega_0=\omega_D=\sqrt{\Omega^2+(\Delta_0')^2}$ for $S$ and $D$ transitions respectively. If $S$ and $D$ are sufficiently apart in $\omega_0$, at $S$, the periodic driving results in Rydberg-blockade despite small $V_0$ [$V_0=0.1\Omega$ in Fig.\ref{fig:1}(a)]. This {\em blockade enhancement} (BE) becomes more apparent later when we analyze HFL or FPL with $\Delta_0=0$. Note that, the blockade exists only if $V_0>\Omega$ for $\delta=\Delta_0=0$. 

Increasing $V_0$, and when $V_0\sim\Omega_{eff}$, the non-interacting picture fails, results a jump in $\omega_D$ to a higher value [Fig.\ref{fig:1}(b)]. Note that, here, $\omega_S$ and $\omega_D$ are provided by the difference in the eigenvalues of the non-driven $\hat H$ ($\delta=0$). The eigenstates can be approximated to $|gg\rangle$, $|+\rangle$ and $|ee\rangle$ if $\Delta_0\gg\Omega$, except when $V_0\approx \Delta_0$ and $V_0\approx 2\Delta_0$ [shaded regions in Fig. \ref{fig:1}(b)-(c)]. Obtaining the eigenvalues up to fourth order in $\Omega$, we have
\begin{eqnarray}
\label{rd1}
\omega_S&=&|E_{gg}-E_+|\approx\Delta_0+\frac{\Omega^2}{2}\left(\frac{2}{\Delta_0}-\frac{1}{\Delta_0-V_0}\right)+\mathcal O\left(\Omega^4\right) \\
\omega_D&=&|E_{gg}-E_{ee}|\approx|V_0-2\Delta_0-\frac{\Omega^2}{2}\left(\frac{2}{\Delta_0}+\frac{1}{\Delta_0-V_0}\right) \nonumber \\
&&+\frac{\Omega^4}{4}\left(\frac{1}{\Delta_0^3}+\frac{1}{(\Delta_0-V_0)^3}\right)|.
\label{rd2}
\end{eqnarray}
Eqs. (\ref{rd1}) and (\ref{rd2}) [dashed and solid lines in Fig. \ref{fig:1}(c)] are in excellent agreement with the numerical results. The jump in $\omega_D$ near $V_0\sim \Omega_{eff}$ is not an abrupt one, but with a small coexistence region. When $V_0\approx\Delta_0$ both $|+\rangle$ and $|ee\rangle$ are almost degenerate, and the resonances $S$ and $D$ as such do not exist, since the population  is shared among both $|+\rangle$ and $|ee\rangle$ from $|gg\rangle$. For large values of $V_0$, $\omega_s$ becomes independent of $V_0$, whereas $\omega_D$ decreases and vanishes at $V_0\simeq2\Delta_0$ and then increases linearly with $V_0$. When $V_0\simeq2\Delta_0$, $|gg\rangle$ and $|ee\rangle$ are almost degenerate, hence $D$ resonance exists even for $\delta=0$, but not $S$. The two resonances cross near $V_0\sim 3\Delta_0+3\Omega^2/\Delta_0$. The resonance widths $w_R$ are obtained by a Lorentzian fit [Fig.\ref{fig:1}(c)]. As $V_0$ increases, $D$ gets progressively narrower making the higher order terms in Eq. (\ref{rd2}) relevant, and its existence at  large $V_0(>2\Delta_0)$ is interpreted as the anti-blockade.

\begin{figure}[hbt]
\centering
\includegraphics[width= 1.\columnwidth]{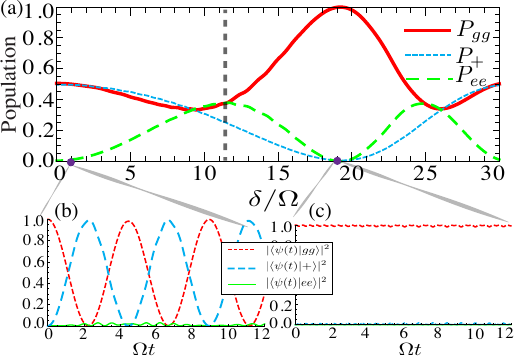}
\caption{\small{(color online).  (a) $P_\alpha$ vs $\delta$ for $N=2$ in the HFL with $n_1=0$, $n_2=-1$ ($\omega_0=V_0=8\Omega$) and $\Omega T_f=100$. At the vertical dashed line ($\delta=11.476 \Omega$) $J_0(\alpha)\sim J_{-1}(\alpha)$. (b) and (c) show the dynamics for $\delta=\Omega$ and $\delta=19.24 \Omega$ respectively.}}
\label{fig:2} 
\end{figure}
 {\em HFL}. The dynamics is better understood by writing the Hamiltonian in Eq. (\ref{hamro}) for $N=2$ as \cite{supp}
  \begin{eqnarray}
 \hat H'^{(2)}&=&\frac{\Omega}{2}\sum_{m-\infty}^{\infty}i^m J_m(\alpha)g(t)\left(\sum_{j=1}^2\hat\sigma_{eg}^j+ \right. \nonumber \\ 
  && \left. \left(\hat\sigma_{eg}^1\hat\sigma_{ee}^2+\hat\sigma_{eg}^2\hat\sigma_{ee}^1\right)\left(e^{iV_0t}-1\right)\right)+ {\rm H.c.} .
 \label{ham2}
 \end{eqnarray}
 A close inspection on Eq. (\ref{ham2}) reveals that $\hat H'$ consists of only off-diagonal elements correspond to $|gg\rangle\leftrightarrow|+\rangle$ and $|+\rangle\leftrightarrow|ee\rangle$ transitions, with respective resonance criteria: (i) $n_1\omega_0=\Delta_0$ and (ii) $n_2\omega_0=\Delta_0-V_0$. Taking large $V_0$ such that the two resonances do not overlap in $\omega_0$, the {\em quantum interference} plays an important role. If condition (i) is met, the transition amplitudes for $|gg\rangle\leftrightarrow|+\rangle$ interfere constructively ($S$ resonance), whereas that of $|+\rangle\leftrightarrow|ee\rangle$ interfere destructively. The opposite is true if (ii) is satisfied. Thus, the dynamics depends crucially on the initial state. For instance, if $|I\rangle=|ee\rangle$, the condition (ii) leads to the coherent ROs between $|ee\rangle$ and $|+\rangle$ with $\Omega'=2\Omega J_{n_2}(\alpha)$, where as PT takes place if $|I\rangle=|gg\rangle$. The state dependent PT emerges as a unique feature from the Rydberg interactions. To satisfy (i) and (ii) simultaneously, we require $n_1\neq n_2$, which leaves $\delta$ as the only free parameter. In Fig. \ref{fig:2}(a) we show the dynamics vs $\delta$ for $n_1=0$, $n_2=-1$, with $V_0=8\Omega$, $\omega_0=8\Omega$ and $|I\rangle=|gg\rangle$. At $\delta=0$ there exists Rydberg blockade, and also the blockade occurs whenever $J_{n_1}(\alpha)\gg J_{n_2}(\alpha)$ for $\delta\neq 0$ with $\Omega'=2\Omega\sqrt{J_0^2(\alpha)+J_{-1}^2(\alpha)}$ [Fig. \ref{fig:2}(b)]. If $J_0(\alpha)\sim J_{-1}(\alpha)$ the anti-blockade occurs, shown by dashed vertical line in Fig. \ref{fig:2}(a). And, PT is shown at $J_0(\alpha)\sim 0$ [Fig. \ref{fig:2}(c)].  

\begin{figure}[hbt]
\centering
\includegraphics[width= 1.\columnwidth]{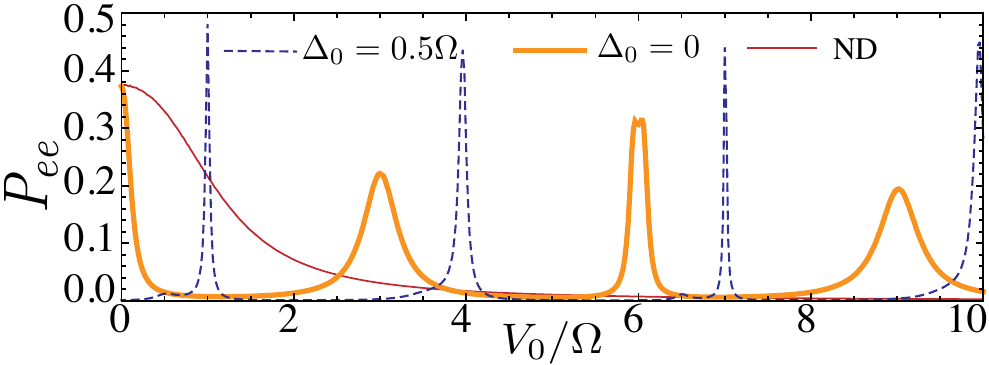}
\caption{\small{(color online). $P_{ee}$ vs $V_0$ in FPL, with $\Omega T_f=300$, $\omega_0=3\Omega$ and $\delta=34 \Omega$. Thin and thick solid lines are for non-driven  and driven cases respectively with $\Delta_0=0$. The dashed line is for $\Delta_0=0.5\Omega$.}}
\label{fig:3} 
\end{figure}

 If $\delta\gtrsim \Omega$ the HFL merges with FPL \cite{ash} for $\Delta_0=0$. The existence of multi-LZTs makes the TM method cumbersome, but for $V_0\gg\Omega_{eff}$, they are well separated in $\Delta_0$ axis. This allows us to separate the adiabatic and non-adiabatic regions and obtain the respective resonance criteria \cite{supp}. Doing so, we get $\Delta_0=n\omega_0$ for $S$ resonance with $\Omega'=(2\omega_0/\pi)|\cos(\theta-\pi/4)|\sqrt{\frac{\pi\Omega^2}{\omega_0}/\sqrt{\delta^2-\Delta_0^2}}$. The resonance condition for $|+\rangle\leftrightarrow |ee\rangle$ is $\Delta_0-V_0=n\omega_0$ with $\Omega'=(2\omega_0/\pi)|\cos(\tilde \theta-\pi/4)|\sqrt{\frac{\pi\Omega^2}{\omega_0}/\sqrt{\delta^2-(\Delta_0-V_0)^2}}$ where $\omega_0\tilde \theta=\sqrt{\delta^2-(\Delta_0-V_0)^2}-(\Delta_0-V_0)\cos^{-1}\frac{\Delta_0-V_0}{\delta}$ and that of $D$ resonance is $\Delta_0-V_0/2=n\omega_0$. They are in agreement with the numerical solutions of Eq. (\ref{ham}). Fig. \ref{fig:3} shows $P_{ee}$ vs $V_0$ in the FPL with $|I\rangle=|gg\rangle$ and $\Delta_0=0$. When $\delta=0$, $P_{ee}$ decreases monotonously (thin line) exhibiting the Rydberg blockade ($P_{ee}\sim 0$) at large $V_0$, whereas in the presence of driving (thick line) it exhibits a non-monotonous character. The initial faster decay of $P_{ee}$ indicates the BE and the periodic peaks at higher $V_0$ show anti-blockades. The peaks can be shifted in $V_0$ as well as made higher or narrower by taking $\Delta_0\neq 0$ (dashed line). The BE at small $V_0$ and anti-blockades at large $V_0$ may have far reaching consequences in the dynamics of periodically driven Rydberg ensembles. Also, we have verified that the two features persists in the presence of spontaneous emission \cite{supp}. 
 
\begin{figure}[hbt]
\centering
\includegraphics[width= 1.0\columnwidth]{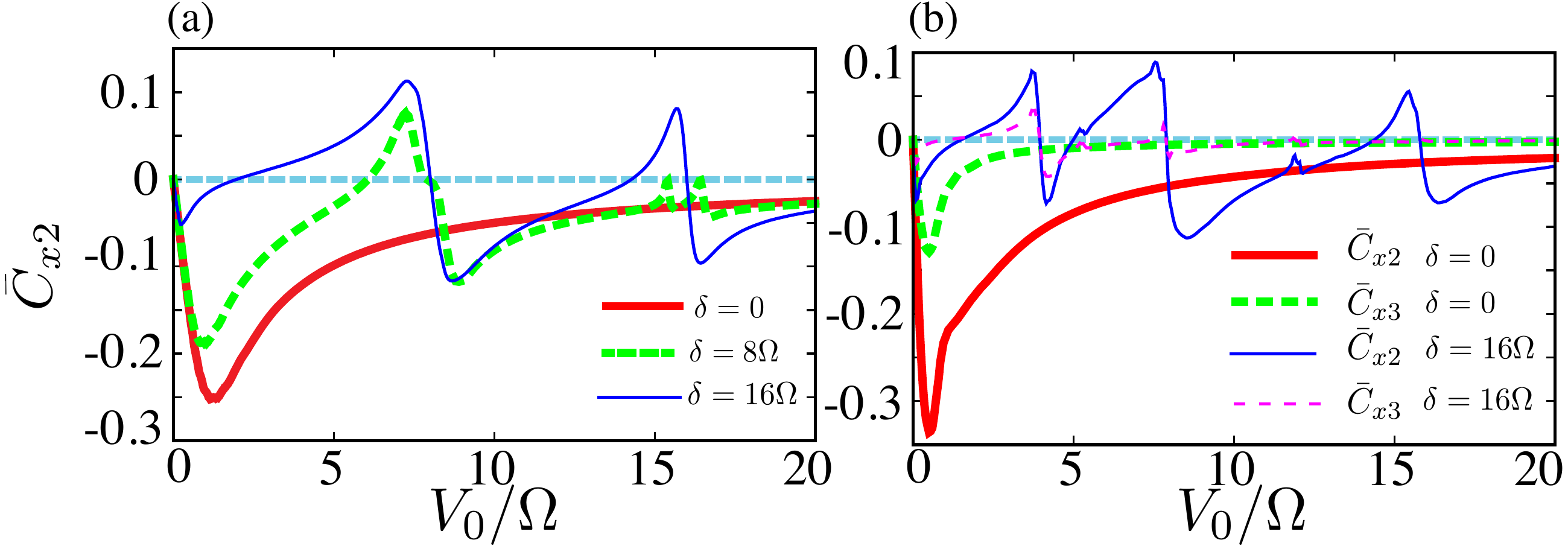}
\caption{\small{(color online). Correlations vs $V_0$ for (a) $N=2$ and (b) $N=10$. For both figures $\Delta_0=0$, $\omega_0=8\Omega$ and $\Omega T_f=150$. }}
\label{fig:4} 
\end{figure}

 {\em CRC}. Interestingly, the last two terms in Eq. (\ref{ham2}) reveal CRC as a fundamental process in Rydberg gases, emerging from the long-range interactions, essentially driving $|+\rangle\leftrightarrow|ee\rangle$ transition. It is also apparent in the period-averaged or the zeroth order Floquet Hamiltonian \cite{rigo14, supp} $H_{eff}=1/T\int_0^Tdt \ \hat H'^{(2)}(t)$, where $T=2\pi/\omega_0$ if $\omega_0\gg V_0$ or $T=2\pi/V_0$ if $V_0\gg \omega_0$. Henceforth, we take $\Delta_0=0$ ($S$-resonance). For $\omega_0\gg V_0$: 
\begin{eqnarray}
\hat H_{eff}^{\omega_0\gg V_0}&=&\frac{\Omega}{2iT} \sum_{m=-\infty}^{\infty}i^mJ_m(\alpha)\frac{(e^{iV_0T}-1)}{m\omega_0+V_0}\hat X \nonumber \\
&&+\frac{J_0(\alpha)\Omega}{2}\left(\sum_{j=1}^2\hat\sigma_{eg}^{j}-\hat X\right)+{\rm H.c.},
\label{te1}
\end{eqnarray}
where $\hat X=\hat\sigma_{ee}^1\hat\sigma_{eg}^2+\hat\sigma_{ee}^2\hat\sigma_{eg}^1$. For $V_0\ll\Omega, \omega_0$, we take $m\omega_0+V_0\approx m\omega_0$ in Eq. (\ref{te1}), which lead us to $\hat H_{eff}^{\omega_0\gg V_0}\simeq J_0(\alpha)\Omega/2\sum_j\hat\sigma_{eg}^j+i[\Omega J_0(\alpha)V_0T/4]\hat X+\mathcal O(V_0^2)+{\rm H.c.}$. Hence, for small interactions the CRC increases linearly with $V_0$. Similarly, for $V_0\gg \omega_0$:
\begin{eqnarray}
\hat H_{eff}^{V_0\gg\omega_0}&=&\frac{\Omega}{2iT} \sum_{m=-\infty}^{\infty}i^mJ_m(\alpha)\left(e^{im\omega_0T}-1\right)\left[\sum_{j=1}^2\frac{\hat\sigma_{eg}^{j}}{m\omega_0} \right. \nonumber \\
&&\left. +\left(\frac{1}{m\omega_0+V_0}-\frac{1}{m\omega_0}\right)\hat X\right]+{\rm H.c.}
\label{te2}
\end{eqnarray}
Eqs. (\ref{te1}) and (\ref{te2}) govern the time evolution of the system at integer multiple of the period $T$. For small amplitude modulations ($\delta\ll 1$), only $m=0, \pm 1$ have significant contributions in Eq. (\ref{te2}), and we get  $\hat H_{eff}^{V_0\gg\omega_0}\simeq \chi(\sum_j\hat\sigma_{eg}^j-\hat X)+{\rm H. c.}$ with $\chi=\Omega[J_0(\alpha)+2iJ_1(\alpha)]/2$, interestingly, which provides an alternative perspective for Rydberg blockade. The -ve sign infront of $\hat X$ implies that {\em it is the correlated Rabi coupling which results in blockade at large $V_0$, by completely suppressing the single atom Rabi coupling} thereby we have $\langle+|\hat H_{eff}^{V_0\gg\omega_0}|ee\rangle=0$. Further, we look at the two-body correlation, $\bar C_{x2}=1/T_f\int_0^{T_f}C_{x2}(t)dt$ with $C_{x2}(t)=\langle\psi(t)|(\hat\sigma_{x}^1\hat\sigma_{ee}^2+\hat\sigma_{x}^2\hat\sigma_{ee}^1)|\psi(t)\rangle/2$ with $|I\rangle=|gg\rangle$ as a function of $V_0$, see Fig. \ref{fig:4}(a).  Its magnitude measures the probability of finding second atom in the Rydberg state while first atom making a transition. When $\delta=0$,  for small $V_0$ the $\bar C_{x2}$ increases with $V_0$ until it reaches a maximum and then decays as $1/V_0$ due to the Rydberg blockade. The periodic modulation ($\delta\neq 0$) results in a non-trivial behaviour for $\bar C_{x2}$, particularly the non-periodic oscillations between +ve and -ve values. Since $\Delta_0=0$, at $V_0=n\omega_0$ both $|+\rangle\leftrightarrow|ee\rangle$ and $|+\rangle\leftrightarrow|gg\rangle$ transitions are at resonance which is identical to the case of $V_0=0$, hence effectively $\bar C_{x2}=0$. Also $\bar C_{x2}$ vanishes when $P_{ee}=0$ making its oscillatory nature. 

$N>2$. More resonances appear with increasing $N$ \cite{c2}. The $S$ resonance ($|G\rangle\leftrightarrow |+\rangle$) is independent of $N$ and $V_0$, but a large $V_0$ is required to isolate it from other resonances. Higher-order correlations,  $C_{x2}(t)=\langle\sum_j(\hat\sigma_x^j\hat\sigma_{ee}^{j+1}+\hat\sigma_x^j\hat\sigma_{ee}^{j-1})\rangle/N $ and $C_{x3}(t)=\langle\sum_j(\hat\sigma_x^j\hat\sigma_{ee}^{j+1}\hat\sigma_{ee}^{j-1})\rangle/N$ for $N=10$ are shown in Fig.\ref{fig:4}(b). As expected $\bar C_{x3}$ is smaller and decays faster with $V_0$ compared to $\bar C_{x2}$. As $N$ increases, the correlations exhibit additional oscillations due to the participation of more resonances.

\begin{figure}[hbt]
\centering
\includegraphics[width= 1.\columnwidth]{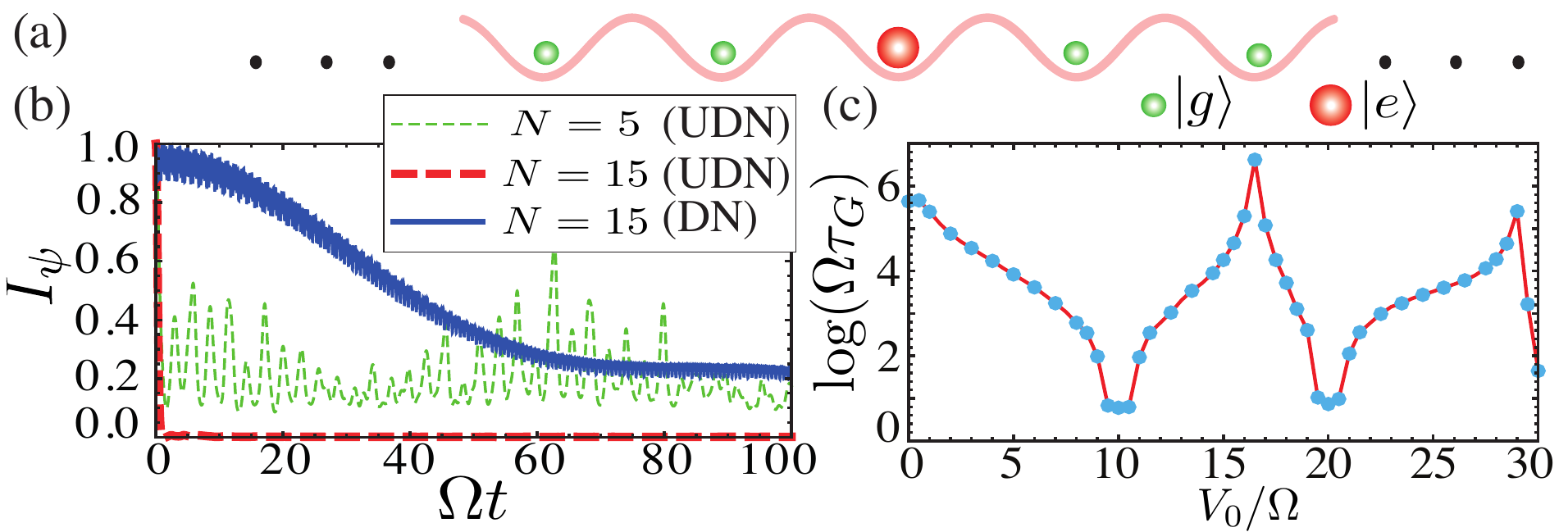}
\caption{\small{(color online). The results for dynamical stabilization. (a) The initial configuration with one excitation at the centre of the lattice. (b) $I_{\psi}$ vs time for both driven (DR) and undriven (UDR) cases with $V_0=5\Omega$. For driven case, $\delta=24.0483 \Omega$ and $\omega_0=10 \Omega$ which give $J_0(\delta/\omega_0)=0$. (c) The log of the temporal width $\tau_G$ vs $V_0$ for $N=15$.}}
\label{fig:5} 
\end{figure}

 In a similar vein, where the dynamical localization of a condensate in a periodically shaken lattice by suppressing the tunnelling is observed \cite{bec2}, we analyze that of a many-body configuration. We take $|I\rangle=|... g, e,  g, ...\rangle$, a singly excited state as depicted in Fig. \ref{fig:5}(a). The localization of $|I\rangle$ in a given eigen-basis can be measured in terms of either survival probability, $|\langle I|\psi(t)\rangle|^2$ \cite{spr} or the inverse participation ratio, $I_{\psi}(t)=\sum_ip_i^2(t)$ \cite{ipr}, where $p_i(t)$ is the probability of finding the system in the $i^{th}$ eigen state. We choose the eigen states of $\hat H (\Omega=0, \delta=0)$ as the basis for our calculations. In the absence of periodic modulation, for small $N$  [dashed line in Fig. \ref{fig:5}(b)] one observes collapse and partial revivals of $|I\rangle$ \cite{zei17}. As $N$ increases, the collapse becomes faster and eventually no revival due to the exponential growth in the dimensions of the Hilbert space. In contrast, the periodic modulation may significantly slow down the collapse, leading to the dynamical localization or stabilization of $|I\rangle$.  

 For $\delta\neq 0$, $I_{\psi}$ exhibits a Gaussian decay \cite{c3} in $t$ [the solid line in Fig. \ref{fig:5}(b)], and the oscillation in the profile is attributed to the LZS interference. Strikingly, similar Gaussian decay of initial state is shown in the quench dynamics of isolated quantum Hamiltonians, when $|I\rangle$ has a Gaussian distribution over the eigen states of the final Hamiltonian \cite{spr}. The Gaussian width $\tau_G$ is found to be independent of $N$ for sufficiently large value of $N$ ($> 5$), but depends crucially on $V_0$ [Fig. \ref{fig:5}(c) shows $\log_{10} \tau_G$ vs $V_0$ for $N=15$] as well as the driving parameters. Since $\Delta_0=0$, choosing $J_0(\delta/\omega_0)=0$ suppress any population transfer to $|G\rangle$ from $|I\rangle$. The transitions to states with $N_e>1$ strongly depends on $V_0$. At $V_0=n\omega$, there is a resonant transition from $|I\rangle$ to doubly excited states with excitations at the nearest sites causing the minima in $\tau_G$. Between those minima, $\tau_G$ acquires a maximum due to the Blockade effect at large interactions.
 
{\em Experimental Parameters}. Taking $\Omega=2\pi\times 1$ MHz, our studies involve $\delta\sim 2\pi\times 0-40$MHz, $\omega_0\sim 2\pi \times 0-15$ MHz, over a maximum time, $T_f \sim 100-300 \mu$s in the HFL or FPL. Note that for $n\sim 80$, the Rubidium $nS$ state has a life time of $600\mu s$ (or the decay constant $\Gamma=2\pi \times 0.00167\Omega$) \cite{bet}. In the supplemental material, we show the results for BE and anti-blockade at large $V_0$ for $N=2$ and $\Gamma=0.01 \Omega$, \cite{supp} which corresponds to $45 S$ state of Rubidium if $\Omega=1$ MHz. 

{\em Conclusions and outlook.} Driving the detuning periodically relaxes the requirements to observe Rydberg Blockade and anti-blockade, thereby acquiring a huge controllability over the quantum dynamics in Rydberg atomic lattices. Our analysis reveals CRC as a novel feature in Rydberg chains, which can be extended as a general characteristic of two level systems with long-range interactions. Further, CRC provides an alternative and a different depiction of Rydberg blockade. 

Our work offers an extra dimension to the problems that can be addressed using Rydberg atomic chains. In particular, the localization of a many-body state addressed in our studies can be extended to analyze ergodic-nonergodic transitions, in the presence of disorder. In otherwords, how heating takes place in such systems under periodic forcing, especially the role of long-range interactions can be probed. The stability (growth and melting) of Rydberg crystals \cite{rycr} under periodic modulation would also be a question of immediate interest due to the state of the art experiments.

{\em Acknowledgemts}:- We acknowledge fruitful discussions with Weibin Li, Vijay Shenoy, Igor Lesanovsky and especially, Anatoli Polkovnikov. R.N. acknowledges the funding by the Indo-French Centre for the Promotion of Advanced Research (CEFIPRA) and support from the UKIERI-UGC Thematic Partnership No. IND/CONT/G/16-17/73UKIERI-UGC project. S.B. acknowledges the support from the Department of Science and Technology (DST), Government of India through the INSPIRE SHE Programme (8S/2012).

\begin{thebibliography}{99}
\bibitem{c3} Similar Gaussian decay is obtained for the fidelity $|\langle I|\psi(t)\rangle|^2$.
\end{thebibliography}

\newpage
\onecolumngrid
\newpage
{
  \center \bf \large 
  Supplemental Material for: \\
  Periodically Driven Array of Single Rydberg Atoms
  \vspace*{0.1cm}\\ 
  \vspace*{0.0cm}
}
\begin{center}
Sagarika Basak, Yashwant Chougale and Rejish Nath\\
  \vspace*{0.15cm}
  \small{\textit{Indian Institute of Science Education and Research, Pune 411 008, India}}\\
  \vspace*{0.25cm}
\end{center}
\twocolumngrid
\section{Hamiltonian in the Rotating frame: Schrieffer-Wolf Transformation}
Introducing Schrieffer-Wolf Transformation defined by the unitary operator, $\hat U(t)=\exp[if(t)\sum_j\hat\sigma_{ee}^{j}+it\sum_{j<k}V_{jk}\hat\sigma_{ee}^{j}\hat\sigma_{ee}^{k}]$ with $f(t)=\delta/\omega_0\cos\omega_0 t-\Delta_0 t$, the Hamiltonian
\begin{equation}
\hat H=-\Delta(t)\sum_{i=1}^N\hat\sigma_{ee}^{i}+\frac{\Omega}{2}\sum_{i=1}^N(\hat\sigma_{eg}^{i}+\hat\sigma_{ge}^{i})+\sum_{i<j}V_{ij}\hat\sigma_{ee}^{i}\hat\sigma_{ee}^{j},
\end{equation}
transforms as $\hat H'(t)=\hat U(t)\hat H(t)\hat U^{\dagger}(t)-i\hbar\hat U(t)\dot{\hat U}^{\dagger}(t)$. The final Hamiltonian is 
\begin{equation}
\hat H'(t)=\frac{\Omega}{2}\hat U(t)\sum_{i=1}^N(\hat\sigma_{eg}^{i}+\hat\sigma_{ge}^{i})\hat U^{\dagger}(t).
\end{equation}
To evaluate $\hat H'(t)$, we need the following terms:
\begin{equation}
 e^{if(t) \sigma_{ee}^i  } \left(\sigma_{eg}^i+\sigma_{ge}^i\right)  e^{-if(t) \sigma_{ee}^i  } = e^{if(t)} \sigma_{eg}^i+e^{-if(t)} \sigma_{ge}^i
\end{equation}
and 
\begin{widetext}
\begin{equation}
e^{it/2 \sum_{j,k}^{j\neq k} V_{jk} \hat\sigma_{ee}^j \hat\sigma_{ee}^k} \left( e^{if(t)}\hat\sigma_{eg}^l+e^{-if(t)}\hat\sigma_{ge}^l\right) e^{-it/2 \sum_{j,k}^{j\neq k} V_{jk} \hat\sigma_{ee}^j \hat\sigma_{ee}^k}  = e^{it \sum_{j}^{j\neq l} V_{jl} \hat\sigma_{ee}^j \hat\sigma_{ee}^l}  \left(e^{if(t)} \hat\sigma_{eg}^l+e^{-if(t)}\hat\sigma_{ge}^l\right)   e^{-it \sum_{j}^{j\neq l} V_{jl} \hat\sigma_{ee}^j \hat\sigma_{ee}^l}, 
\end{equation}
\end{widetext}
where the double summation in the exponential function has reduced to a single one in the last step. Then, using Baker-Hausdorff lemma we get,
\begin{eqnarray}
e^{it \sum_{j}^{j\neq l} V_{jl} \hat\sigma_{ee}^j \hat\sigma_{ee}^l}  \left(e^{if(t)} \hat\sigma_{eg}^l+e^{-if(t)}\hat\sigma_{ge}^l\right)   e^{-it \sum_{j}^{j\neq l} V_{jl} \hat\sigma_{ee}^j \hat\sigma_{ee}^l} \nonumber \\
= e^{i\left[f(t)+t\sum_{j\neq l}V_{lj}\hat\sigma_{ee}^{j}\right]}\hat\sigma_{eg}^l+e^{-i\left[f(t)+t\sum_{j\neq l}V_{lj}\hat\sigma_{ee}^{j}\right]}\hat\sigma_{ge}^l,
\end{eqnarray}
which then finally gives us,
\begin{eqnarray}
\hat H'&=&\frac{\Omega}{2}\sum_{j=1}^N\sum_{m=-\infty}^{\infty}i^mJ_m(\alpha)e^{i(m\omega_0-\Delta_0+\sum_{k\neq j}V_{jk}\hat\sigma_{ee}^{k})t}\hat\sigma_{eg}^{j}+{\rm H.c.}  \nonumber \\
\end{eqnarray}
where $J_m(\alpha)$ is the $m$th order Bessel function with $\alpha=\delta/\omega_0$. Using $e^{\pm i\sum_{k\neq j}V_{jk}\hat\sigma_{ee}^{k}t}=\prod_{k\neq j}  \left[\hat\sigma_{ee}^k (e^{\pm itV_{jk}}-1)+\mathcal I\right]$, where $\mathcal I$ is the identity operator, we can rewrite the Hamiltonian as 
\begin{eqnarray}
\hat H'(t)&=& \frac{\Omega}{2}\sum_{j}^{N} \sum_{m=-\infty}^{\infty}i^mJ_m(\alpha)\left[g(t)\hat\sigma_{eg}^j\left(\prod_{k\neq j} \left[\hat\sigma_{ee}^k (e^{itV_{jk}}-1)+\mathcal I \right]  \right) \right.\nonumber\\
&& \left. + g^*(t) \hat\sigma_{ge}^j  \left(\prod_{k\neq j} \left[\hat\sigma_{ee}^k (e^{-itV_{jk}}-1)+\mathcal I \right]  \right)\right],
\label{hro2}
\end{eqnarray}
where $g(t)=\exp[i(m\omega_0-\Delta_0)t]$. The Eq. (\ref{hro2}) for $N=2$ is discussed in the main text in detail.
\section{ Nearest neighbour approximation: Time independent Hamiltonian}
We truncate the interactions beyond nearest neighbour, and then calculate the time independent average Hamiltonian, $H_{eff}=1/T\int_0^Tdt \ \hat H'(t)$ where $T=2\pi/\omega_0$ if $\omega_0\gg V_0$ or $T=2\pi/V_0$ if $V_0\gg\omega_0$ . Up to nearest neighbour interactions, Eq. (\ref{hro2}) reduces to
\begin{eqnarray}
\hat H_{NN}'(t)&=& \frac{\Omega}{2}\sum_{j}^{N} \sum_{m=-\infty}^{\infty}i^mJ_m(\alpha)g(t)\hat\sigma_{eg}^j\left[1+\hat\sigma_{ee}^{j+1}\hat\sigma_{ee}^{j-1}\left(e^{iV_0t}-1\right)^2\right.\nonumber\\
 &&\left.+\left(\hat\sigma_{ee}^{j+1}+\hat\sigma_{ee}^{j-1}\right)\left(e^{iV_0t}-1\right) \right]+{\rm H.c.}. \nonumber \\
\label{hronn}
\end{eqnarray}
{\bf Case 1}: The effective time independent Hamiltonian up to the nearest neighbour interaction $V_0$ for $\omega_0\gg V_0$: 
\begin{widetext}
\begin{eqnarray}
\hat H_{eff}^{\omega_0\gg V_0}&=&\frac{\Omega}{2iT}\sum_{j}^{N} \sum_{m=-\infty}^{\infty}i^mJ_m(\alpha)\left[\frac{\left(e^{-i\Delta_0T}-1\right)}{m\omega_0-\Delta_0}\left(1-\hat\sigma_{ee}^{j+1}-\hat\sigma_{ee}^{j-1}+\hat\sigma_{ee}^{j+1}\hat\sigma_{ee}^{j-1}\right)+\frac{\left(e^{-i(\Delta_0-2V_0)T}-1\right)}{m\omega_0-\Delta_0+2V_0}\hat\sigma_{ee}^{j+1}\hat\sigma_{ee}^{j-1}\right.\nonumber \\
&&\left. +\frac{\left(e^{-i(\Delta_0-V_0)T}-1\right)}{m\omega_0-\Delta_0+V_0}\left(\hat\sigma_{ee}^{j+1}+\hat\sigma_{ee}^{j-1}-2\hat\sigma_{ee}^{j+1}\hat\sigma_{ee}^{j-1}\right)\right]\hat\sigma_{eg}^j+{\rm H.c.}.
\end{eqnarray}
\end{widetext}
For $\Delta_0=0$, it becomes
\begin{widetext}
\begin{eqnarray}
\hat H_{eff}^{\omega_0\gg V_0}&=&\frac{\Omega}{2iT}\sum_{j}^{N} \sum_{m=-\infty}^{\infty}i^mJ_m(\alpha)\left[\frac{\left(e^{i2V_0T}-1\right)}{m\omega_0+2V_0}\hat\sigma_{ee}^{j+1}\hat\sigma_{ee}^{j-1}+\frac{\left(e^{iV_0T}-1\right)}{m\omega_0+V_0}\left(\hat\sigma_{ee}^{j+1}+\hat\sigma_{ee}^{j-1}-2\hat\sigma_{ee}^{j+1}\hat\sigma_{ee}^{j-1}\right)\right]\hat\sigma_{eg}^j \nonumber \\
&&+\frac{J_0(\alpha)\Omega}{2}\sum_{j=1}^N\hat\sigma_{eg}^{j}\left(1-\hat\sigma_{ee}^{j+1}-\hat\sigma_{ee}^{j-1}+\hat\sigma_{ee}^{j+1}\hat\sigma_{ee}^{j-1}\right)+{\rm H.c.},
\label{e1}
\end{eqnarray}
\end{widetext}
Now, for $V_0\ll\Omega$, the Eq. (\ref{e1}) becomes
\begin{widetext}
\begin{eqnarray}
\hat H_{eff}^{\omega_0\gg V_0}&=&\frac{\Omega}{2}\sum_{j}^{N} \sum_{m=-\infty}^{\infty}i^mJ_m(\alpha)\left[\frac{-2V_0^2}{(m\omega_0+2V_0)(m\omega_0+V_0)}\hat\sigma_{ee}^{j+1}\hat\sigma_{ee}^{j-1}+\frac{V_0}{m\omega_0+V_0}\left(\hat\sigma_{ee}^{j+1}+\hat\sigma_{ee}^{j-1}\right)\right]\hat\sigma_{eg}^j\nonumber \\
&&+\frac{J_0(\alpha)\Omega}{2}\sum_{j=1}^N\hat\sigma_{eg}^{j}\left(1-\hat\sigma_{ee}^{j+1}-\hat\sigma_{ee}^{j-1}+\hat\sigma_{ee}^{j+1}\hat\sigma_{ee}^{j-1}\right)+{\rm H.c.},
\label{e2}
\end{eqnarray}
\end{widetext}
Writing  Eq. (\ref{e2}) for $N=2$:
\begin{eqnarray}
\hat H_{eff}^{\omega_0\gg V_0}&=&\frac{\Omega}{2} \sum_{m=-\infty}^{\infty}i^mJ_m(\alpha)\left[\frac{V_0}{m\omega_0+V_0}\left(\hat\sigma_{eg}^1\hat\sigma_{ee}^{2}+\hat\sigma_{eg}^2\hat\sigma_{ee}^{1}\right)\right] \nonumber \\
&&+\frac{J_0(\alpha)\Omega}{2}\left(\sum_{j=1}^2\hat\sigma_{eg}^{j}-\hat\sigma_{eg}^1\hat\sigma_{ee}^{2}-\hat\sigma_{eg}^2\hat\sigma_{ee}^{1}\right)+{\rm H.c.}.
\label{e22}
\end{eqnarray}
{\bf Case 2}: Similarly,  for $V_0\gg \omega_0$: 
\begin{widetext}
\begin{eqnarray}
\hat H_{eff}^{V_0\gg\omega_0}=\frac{\Omega}{2iT}\sum_{j}^{N} \sum_{m=-\infty}^{\infty}i^mJ_m(\alpha)\left(g(T)-1\right)\left[\frac{\left(1-\hat\sigma_{ee}^{j+1}-\hat\sigma_{ee}^{j-1}+\hat\sigma_{ee}^{j+1}\hat\sigma_{ee}^{j-1}\right)}{m\omega_0-\Delta_0}+\frac{\hat\sigma_{ee}^{j+1}\hat\sigma_{ee}^{j-1}}{m\omega_0-\Delta_0+2V_0}
 +\frac{\left(\hat\sigma_{ee}^{j+1}+\hat\sigma_{ee}^{j-1}-2\hat\sigma_{ee}^{j+1}\hat\sigma_{ee}^{j-1}\right)}{m\omega_0-\Delta_0+V_0}\right]\hat\sigma_{eg}^j+{\rm H.c.}. \nonumber \\
 \label{eev}
\end{eqnarray}
\end{widetext}
For $\Delta_0=0$, Eq. (\ref{eev}) becomes
\begin{widetext}
\begin{eqnarray}
\hat H_{eff}^{V_0\gg\omega_0}=\frac{\Omega}{2iT}\sum_{j}^{N} \sum_{m=-\infty}^{\infty}i^mJ_m(\alpha)\left(e^{im\omega_0T}-1\right)\left[\frac{\left(1-\hat\sigma_{ee}^{j+1}-\hat\sigma_{ee}^{j-1}+\hat\sigma_{ee}^{j+1}\hat\sigma_{ee}^{j-1}\right)}{m\omega_0}+\frac{\hat\sigma_{ee}^{j+1}\hat\sigma_{ee}^{j-1}}{m\omega_0+2V_0}
 +\frac{\left(\hat\sigma_{ee}^{j+1}+\hat\sigma_{ee}^{j-1}-2\hat\sigma_{ee}^{j+1}\hat\sigma_{ee}^{j-1}\right)}{m\omega_0+V_0}\right]\hat\sigma_{eg}^j+{\rm H.c.}. \nonumber \\
 \label{eev}
\end{eqnarray}
\end{widetext}
For $N=2$, we have 
\begin{eqnarray}
\hat H_{eff}^{V_0\gg\omega_0}&=&\frac{\Omega}{2iT} \sum_{m=-\infty}^{\infty}i^mJ_m(\alpha)\left(e^{im\omega_0T}-1\right)\left[\sum_{j=1}^2\frac{\hat\sigma_{eg}^{j}}{m\omega_0} \right. \nonumber \\
&&\left. +\left(\frac{1}{m\omega_0+V_0}-\frac{1}{m\omega_0}\right)\hat X\right]+{\rm H.c.}
\end{eqnarray}


\section{Transverse matrix approach for two interacting Rydberg atoms}
\begin{figure}[hbt]
\centering
\includegraphics[width= 1.\columnwidth]{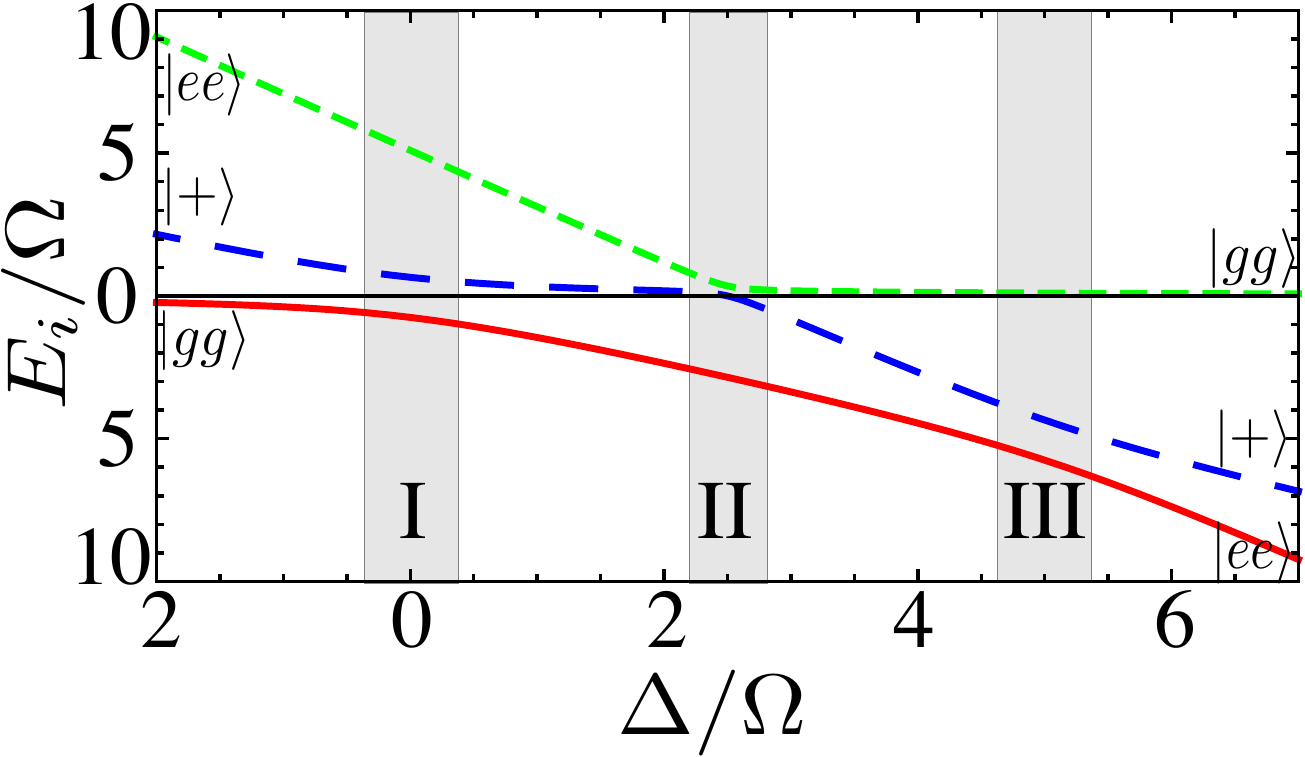}
\caption{\small{(color online). The energy levels vs $\Delta$ obtained by diaganolizing the Hamiltonian in Eq. \ref{ham} with $\delta=0$ and $V_0/\hbar\Omega=5$, in the basis $\{|gg\rangle,|+\rangle, |ee\rangle\}$. The asymptotic states are also shown in the figure. The three shaded regions are the different LZT regions asymptotically corresponds to (I) $|gg\rangle\leftrightarrow |+\rangle$, (II) $|gg\rangle\leftrightarrow |ee\rangle$ and (III) $|+\rangle\leftrightarrow |ee\rangle$.}}
\label{fig:6} 
\end{figure}
The energy level diagram as a function of $\Delta$ in the non-driven case ($\delta=0$) is shown in Fig. \ref{fig:5} for $V_0=5\Omega$. We consider each of the LZT points separately.

\subsection{First transition: $|gg\rangle\leftrightarrow |+\rangle$}
Under strong driving, the LZT matrix written in the basis $\{|gg\rangle,|+\rangle, |ee\rangle\}$ is,
\[
G^{gg \rightarrow +}_{LZ,k}=
\begin{bmatrix}
\cos\chi_1/2 &  -e^{i\theta_{LZ,k}} \sin\chi_1/2 &0\\
e^{-i\theta_{LZ,k}}\sin\chi_1/2 & \cos\chi_1/2&0 \\
0&0&0 \\
\end{bmatrix}
\]
where $k$ is the direction of the sweep across the crossing: for $k = 1$, $\Delta$(t) goes
from +ve  to -ve and viceversa for $k = 2$. The sweep rates are direction independent. Hence, $\chi_1$ is independent of $k$ and $\cos^2\chi_1/2=e^{-\pi \Omega^2/v_1}$, where $v_1=\omega_0\sqrt{ \delta^2 - \Delta_0^2}$ is the rate at which the atom is swept through the avoided level crossing and is obtained by linearizing the $\hat H$ around the LZT point.  The boundary independent phases acquired during the LZT are: $ \hat{\theta}_{LZ,1} = \pi - \phi_S$ and  $\hat{\theta}_{LZ,2} =  \phi_S$, where the Stokes phase, 
\begin{equation}
\phi_{S} = \dfrac{\pi}{4} + \delta^{'}(\ln\delta^{'} -1) + \arg[\Gamma(1-i\delta^{'})],
\end{equation}
with $\Gamma$ is the gamma function and $\delta^{'}=\Omega^2/2v$. $\phi_{S}$ approaches $\dfrac{\pi}{4}$ in the diabatic limit and 0 in the adiabatic limit. Away from the LZT region, the states acquire a relative phase given by the adiabatic matrix
\[
G^{gg \rightarrow +}_{j}=
\begin{bmatrix}
e^{-i \theta_j} & 0& 0\\
0 & e^{i \theta_j}&0 \\
0 &0&1 \\
\end{bmatrix}.
\]
If $\Delta_0$ $\neq$ 0, there are two phase factors corresponding to the system being on the right or left side of the crossing region. Finally, the evolution matrix for one full cycle is $G = G_{LZ,2}^{gg \rightarrow +}G_2G_{LZ,1}^{gg \rightarrow +}G_1$. Rewriting as, $\hat G=\hat G_{xy}\hat G_{z}$ where $\hat G_{xy}$ and $\hat G_{z}$ are respectively represent rotations about an axis lying in the $xy$ plane and $z$ axis \cite{ash}. The former introduces the population transfer between the diabatic states, and the latter is just an overall phase matrix with diagonal elements. In FPL, the overall phase matrix can be approximated as $\{e^{-i(\theta_1+\theta_2)}, e^{i(\theta_1+\theta_2)}, 1\}$. The complete population transfer between the states happens, i.e. the resonance occurs when $\theta_1+\theta_2=2n\pi$ (constructive interference), which then gives us the $S$ resonance condition: $\Delta_0=n\omega_0$.
\subsection{Second transition: $|gg\rangle\leftrightarrow |ee\rangle$}
The population transfer from $|gg\rangle$ to $|ee\rangle$ takes place through the $|+\rangle$ state. Hence, the landau zener matrix is the product of the two landau zener matrices defined for ground to plus and plus to excited state:
\begin{align*}
G^{gg \rightarrow ee}_{LZ,k} =  G^{+ \rightarrow ee}_{LZ,k} \cdot  G^{gg \rightarrow +}_{LZ,k}
\end{align*}
Here, the adiabatic phase matrix is given by, 
\[
G^{gg \rightarrow ee}_{j}=
\begin{bmatrix}
e^{-i \theta_j} &0&0 \\
0&e^{-i(\kappa_j + \theta_j)} & 0\\
0&0 & e^{i \kappa_j} \\
\end{bmatrix}
\]
where $\kappa_j$ is the adiabatic phase acquired away from the $|+\rangle$ to $|ee\rangle$ avoided crossing. Finally, after doing similar analysis like the first transition we arrive at the resonance condition: $\Delta_0-V_0/2=n\omega_0$.

\subsection{Third transition: $|+\rangle\leftrightarrow |ee\rangle$}
The LZT matrix attained for this transition is,
\[
G^{+ \rightarrow ee}_{LZ,k}=
\begin{bmatrix}
0&0&0 \\
0&\cos\chi_2/2 &  -e^{i\theta_{LZ,k}} \sin\chi_2/2\\
0&e^{-i\theta_{LZ,k}}\sin\chi_2/2 & \cos\chi_2/2 \\
\end{bmatrix}
\]
where $\cos^2\chi_2/2 =  e^{-\pi \Omega^2/v_2}$ with the sweep rate at the LZ crossing, $v_2 = \omega_0\sqrt{\delta^2 - (\Delta_0-V)^2}$. The phase matrix attained is, 
\[
G^{+ \rightarrow ee}_{j}=
\begin{bmatrix}
1 &0&0 \\
0&e^{-i\kappa_j} & 0\\
0&0 & e^{i \kappa_j} \\
\end{bmatrix}
\]
A similar analysis as above, gives us a resonance condition: $\Delta_0-V_0=n\omega_0$.
\section{Different Resonances for $N=3$}
There are a total of 9 resonances in a 3-atom lattice. They are:
\newline
\begin{tabular}{ |p{4.5cm}|p{4cm}|}
	\hline
	Transition & Resonance condition\\
	\hline
	$|ggg\rangle \leftrightarrow |gge\rangle, |geg\rangle, |egg\rangle$   & $n\omega_0=\Delta_0$   \\
	$|ggg\rangle \leftrightarrow |gee\rangle, |eeg\rangle$   & $n\omega_0=\Delta_0-V_0/2$ \\
	$|ggg\rangle \leftrightarrow |ege\rangle$   & $n\omega_0=\Delta_0-V_0/128$ \\
	$|ggg\rangle \leftrightarrow |eee\rangle$   & $n\omega_0=\Delta_0-2V_0/3-V_0/192$ \\
	$|gge\rangle, |geg\rangle, |egg\rangle \leftrightarrow  |gee\rangle, |eeg\rangle$  & $n\omega_0=\Delta_0-V_0$   \\
	$|gge\rangle, |geg\rangle, |egg\rangle \leftrightarrow  |ege\rangle$  & $n\omega_0=\Delta_0-V_0/64$   \\
	$|gge\rangle, |geg\rangle, |egg\rangle \leftrightarrow  |eee\rangle$  & $n\omega_0=\Delta_0-V_0-V_0/128$   \\
	$|gee\rangle, |eeg\rangle \leftrightarrow  |eee\rangle$  & $n\omega_0=\Delta_0-V_0-V_0/64$   \\
	$|ege\rangle \leftrightarrow  |eee\rangle$  & $n\omega_0=\Delta_0-2V_0$   \\
	\hline
\end{tabular}
If the interactions are small, all of them overlap with each other. 
\subsection{Dissipative dynamics}
\begin{figure}[hbt]
\centering
\includegraphics[width= 1.\columnwidth]{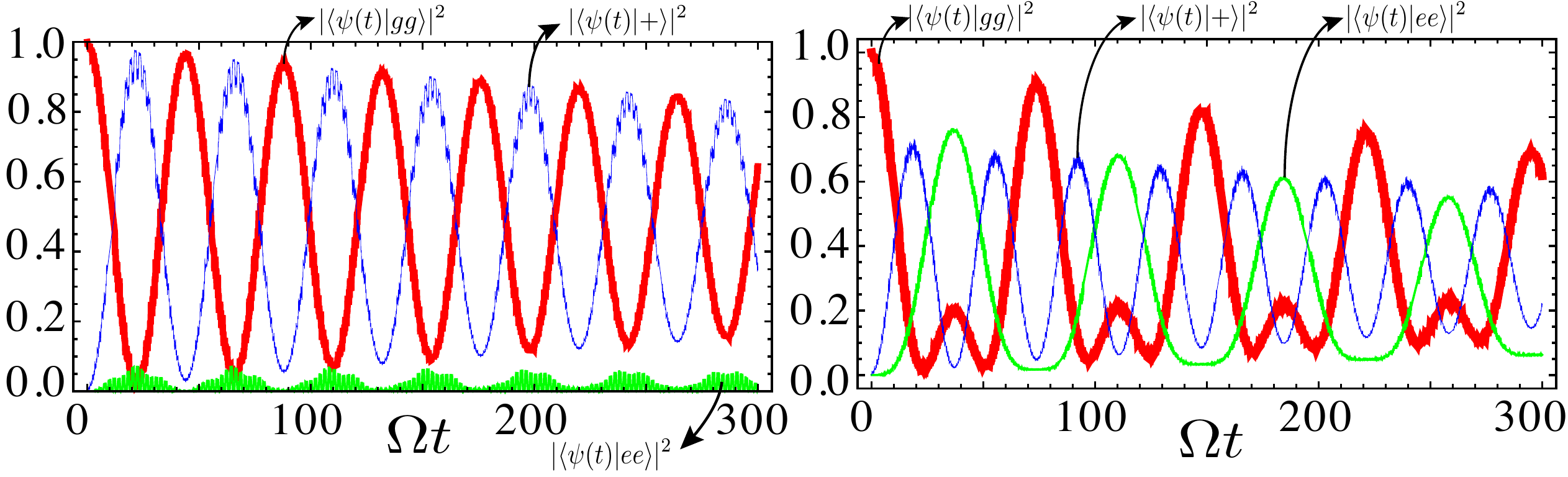}
\caption{\small{(color online). (a) The blockade enhancement at $V_0=0.5\Omega$ and (b) anti-blockade dynamics at $V_0=6\Omega$ for $N=2$, $\omega_0=3\Omega$, $\delta=34\Omega$, $\Delta_0=0$ and $\Omega=1$MHz.  The Rydberg state is taken to be 43$S_{1/2}$ state, which has a life time of $\sim 100 \mu s$.}}
\label{fig:7} 
\end{figure}

To study the effect of spontaneous emission on the two atoms correlated dynamics, we introduce the master equations for the two-particle density matrix,
\begin{equation}
\partial_t \hat{\rho} = -i \left[\hat{H},\hat{\rho}\right]  +\mathcal{L} [\hat{\rho}],
\end{equation}
with the Lindblad operator given by 
\begin{equation}
\mathcal{L}[\rho] =   \sum_{i=1}^2\hat C_i \hat{\rho} \hat C^{\dagger}_i- \frac{1}{2} \sum_m \left(\hat C^{\dagger}_i \hat C_i \hat{\rho} + \hat{\rho}\hat C^{\dagger}_i \hat C_i\right)
\end{equation}
where the operator, $\hat C_i= \sqrt{\Gamma} \hat\sigma_{ge}^i$ with $\Gamma$ is the spontaneous decay rate of the excited state $|e\rangle$. The results for blockade enhancement and anti-blockade at large interactions are shown in Fig. \ref{fig:7} for $N=2$.

 \end{document}